\documentstyle[12pt]{article}
\font\tenrm=cmr10
\font\tenit=cmti10
 1
 1
 1

\def\Re{{\cal R \mskip-4mu \lower.1ex \hbox{\it e}\,}}
\def\Im{{\cal I \mskip-5mu \lower.1ex \hbox{\it m}\,}}

\def\sub#1{_{\lower.25ex\hbox{$\scriptstyle#1$}}}
\def\sul#1{_{\kern-.1em#1}}
\def\sll#1{_{\kern-.2em#1}}
\def\sbl#1{_{\kern-.1em\lower.25ex\hbox{$\scriptstyle#1$}}}
\def\ssb#1{_{\lower.25ex\hbox{$\scriptscriptstyle#1$}}}
\def\sbb#1{_{\lower.4ex\hbox{$\scriptstyle#1$}}}

\def\mh{\ifmmode m\sbl H \else $m\sbl H$\fi}
\def\mch{\ifmmode m_{H^\pm} \else $m_{H^\pm}$\fi}
\def\mt{\ifmmode m_t\else $m_t$\fi}
\def\mc{\ifmmode m_c\else $m_c$\fi}
\def\mz{\ifmmode M_Z\else $M_Z$\fi}
\def\mw{\ifmmode M_W\else $M_W$\fi}
\def\mws{\ifmmode M_W^2 \else $M_W^2$\fi}
\def\mhs{\ifmmode m_H^2 \else $m_H^2$\fi}
\def\mzs{\ifmmode M_Z^2 \else $M_Z^2$\fi}
\def\mts{\ifmmode m_t^2 \else $m_t^2$\fi}
\def\mcs{\ifmmode m_c^2 \else $m_c^2$\fi}
\def\mchs{\ifmmode m_{H^\pm}^2 \else $m_{H^\pm}^2$\fi}
\def\ztwo{\ifmmode Z_2\else $Z_2$\fi}
\def\zone{\ifmmode Z_1\else $Z_1$\fi}
\def\mtwo{\ifmmode M_2\else $M_2$\fi}
\def\mone{\ifmmode M_1\else $M_1$\fi}
\def\tb{\ifmmode \tan\beta \else $\tan\beta$\fi}
\def\xw{\ifmmode x\sub w\else $x\sub w$\fi}
\def\ch{\ifmmode H^\pm \else $H^\pm$\fi}
\def\lum{\ifmmode {\cal L}\else ${\cal L}$\fi}
\def\inpb{\ifmmode {\rm pb}^{-1}\else ${\rm pb}^{-1}$\fi}
\def\infb{\ifmmode {\rm fb}^{-1}\else ${\rm fb}^{-1}$\fi}
\def\epem{\ifmmode e^+e^-\else $e^+e^-$\fi}
\def\ppb{\ifmmode \bar pp\else $\bar pp$\fi}

\newskip\zatskip \zatskip=0pt plus0pt minus0pt
\def\matth{\mathsurround=0pt}

\def\atversim#1#2{\lower0.7ex\vbox{\baselineskip\zatskip\lineskip\zatskip
\lineskiplimit 0pt\ialign{$\matth#1\hfil##\hfil$\crcr#2\crcr\sim\crcr}}}

\renewcommand{\thefootnote}{\fnsymbol{footnote}}

\begin{document} \begin{titlepage}
\rightline{\vbox{\halign{&#\hfil\cr
&PAR/ LPTHE 93 - 05\cr
&February 1993\cr}}}
\vspace{1in}
\begin{center}

{\Large\bf  Possible huge enhancement in the radiative decay of the weak W
boson into the charmed $D_{s} $ meson.}
\footnote  {\em Laboratoire associ\'e au CNRS,URA 280\\
Postal address: Universit\'e Pierre et Marie Curie, Tour 16, $1^{er} $ \'etage,
\\ 4 Place jussieu, 75252 Paris CEDEX 05-France\\
Email :keum@lpthe.jussieu.fr and pham@lpthe.jussieu.fr \\ }

\bigskip

\normalsize
Y. Y. KEUM  and X. Y. PHAM\\

\medskip
{\small \it Laboratoire de Physique Th\'eorique et Hautes Energies, Paris,
France$^{*}$ \\ }

\end{center}
\bigskip
\bigskip

\begin{abstract}

We point out that the rare decay mode $W^{\pm} \rightarrow \gamma + D_{s}^{\pm}
$ could be spectacularly enhanced, with a branching ratio around $10^{-6} $
which is three order of magnitude larger than previous predictions. Its
observation will determine the W boson mass with great accuracy, providing
additional high precision tests of the standard model, as well as reveal
eventual deviation from the trilinear non-abelian gauge coupling.

\end{abstract}

\renewcommand{\thefootnote}{\arabic{footnote}} \end{titlepage}

\begin{center}{{\large \bf   Possible huge enhancement in the radiative decay
of the weak W boson into the charmed $D_{s} $ meson. \\}
\vglue 0.6cm
\normalsize
{\tenrm Y. Y. Keum and X. Y. Pham \\}
\baselineskip=13pt
{\tenit   Laboratoire de Physique Th\'eorique et Hautes Energies, Paris,France
\\}
\baselineskip=12pt
\vglue 0.2cm
{\tenrm ABSTRACT}}

\end{center}
\vglue 0.05cm
{\rightskip=3pc
 \leftskip=3pc
 \tenrm\baselineskip=12pt
 \noindent
We point out that the rare decay mode $W^{\pm} \rightarrow \gamma + D_{s}^{\pm}
$ could be spectacularly enhanced, with a branching ratio around $10^{-6} $
which is three order of magnitude larger than previous predictions. Its
observation will determine the W boson mass with great accuracy, providing
additional high precision tests of the standard model, as well as reveal
eventual deviation from the trilinear non-abelian gauge coupling.
\vglue 0.6cm}
\tenrm
\hspace{35mm}..........................................................\\

\normalsize
In this paper, we present a study of the exclusive radiative decay mode of the
weak charged boson $W^{\pm} $ into a photon and a pseudoscalar meson
$D_{s}^{\pm}(1969) $, which is a  $c\overline{s} $ bound state of  charm and
strange quarks. The branching ratio $B(W^{\pm} \rightarrow \gamma +
D_{s}^{\pm}) $ we obtain is about $10^{-6} $ which is thousand times larger
than previous estimates by Arnellos, Marciano and Parsa $^{[1]} $, referred as
AMP in the following. Matched to the inclusive radiative decay $W^{+}
\rightarrow \gamma + c + \overline{s} $ for which the branching ratio is
calculated $^{[2]} $ to be around $5 \cdot 10^{-4} $, our result suggests that
the probability for a pair of $c + \overline{s} $ quarks to form a pseudoscalar
$D_{s}^{+} $ is few per mille, which is quite plausible. For such radiative (
and rare ) decay mode, this branching ratio is large and could be observable
with the increase in statistics in the future. If detected, this exclusive
process provides useful informations regard

ing electroweak and strong dynamics, its interests lie in the following reasons
:

\hspace{10mm} 1. The trilinear non-abelian gauge coupling, manifestly involved
in this decay (Fig.1), and its eventual deviation from the Yang-Mills structure
can be tested.

\hspace{10mm} 2. Detection of the two-body decay mode $W^{\pm} \rightarrow
\gamma + D_{s}^{\pm} $, combined with the measurement of the photon energy,
provides a very precise determination of the W boson mass which is still poorly
known within ${\pm} 260 $ MeV  errors at the present time ( to be compared with
${\pm} 7 $ MeV for the $Z^{0} $ ). The accuracy value of $M_{W} $ obtained from
this mode will represent a big jump in the high precision test program of the
Standard Model.

\hspace{10mm} 3. In the triangular diagrams ( Fig 2, 3 ) that we employ to
compute the rate ( what could be else ? ), an effective Yukawa coupling
$\lambda    $ between the pseudoscalar meson P and the quark pair
$\overline{q}q $ is introduced via the interaction
$\lambda\overline{q}\gamma_{5}qP $, and $\lambda $ must be determined. Using,
as a guide, the pion-up quark effective coupling
$\lambda_{\overline{u}u\pi^{o}} = \sqrt{2} m_{u}/ F_{\pi} $ which is the
Goldberger-Treiman ( G-T ) relation extended to the quark level$^{[F1]} $, and
\hspace{30mm}.................................................................\\ \small \em
\hspace{3mm}[F 1]. We note that the coupling $\lambda_{\overline{u}u\pi^{o}}$
is in perfect agreement with the Adler-Bell-Jackiw anomaly$^{[3,4]} $ and can
be derived within the Nambu, Jona-Lasinio model$^{[5,6]} $.

\vspace{5mm}
\normalsize \rm
\noindent
will be discussed later in details, then the mode $ W^{\pm} \rightarrow \gamma
+ D_{s}^{\pm} $ could be used to test our model (i.e. the triangular diagrams
together with the  extended G-T relation ) and consequentely the dynamics of
hadronization from quarks.

\hspace{3mm} The total decay amplitude is described by the three Feynman
 diagrams of Figs 1, 2 and 3 :

\vspace{60mm}

\hspace{6mm} Fig.1 \hspace{40mm} Fig.2 \hspace{40mm} Fig.3 \\
\vspace{5mm}

\hspace{3mm} The first one ( Fig. 1 ) can be written as :
\begin{displaymath}
 Q_{1}  =  -i(eg / 2 \sqrt{2})  V_{cs} \varepsilon^{\mu}(P)
\overline{\varepsilon}^{\nu}(\it k) \{ g_{\mu\nu}(P +\it k)_{\alpha} - 2
g_{\nu\alpha} \it k_{\mu} - 2 g_{\mu\alpha} P_{\nu} \}
\end{displaymath}
\begin{equation}
\quad
       \{( - g^{\alpha\beta} + \frac{(P - \it k)^{\alpha}(P - \it
k)^{\beta}}{M_W^2} \ )  / ((P - \it k)^2 - M_W^2) \} \{i F_D(P - k)_{\beta} \}
\;.
\label{eq:1}
\end{equation}
In Eq.(1), the conditions $P_{\mu}\varepsilon^{\mu}(P) = \it
k_{\nu}\varepsilon^{\nu}(\it k) = $ 0  being taken account, $V_{cs} $ is the
Cabibbo-Kobayashi-Maskawa charm-strange transition, $F_{D} $ is the $D_{s} $
decay constant ( analogous to the pion one : $F_{\pi} \simeq 132 $ MeV )
involved in $D_{s} \rightarrow \tau \nu $ for example and $g = e/sin\theta_{W}
$ is the SU(2) gauge coupling.\\

Contracting the indices $\alpha$ and $\beta$, the amplitude $Q_{1}$ reduces to
a very simple expression $^{[1]} $ :
\begin{equation}
 Q_{1}  = (eg /2 \sqrt{2}) V_{cs} F_{D} \varepsilon^{\mu}(P)
\overline{\varepsilon}^{\nu}(\it k) g_{\mu\nu} \;.
\label{eq:2}
\end{equation}

We remark that this reduction holds only if the tri-linear coupling obeys the
standard non-abelian structure as explicited in Eq.(1).

For the diagrams of Figs(2) and (3), let us denote by $\lambda_{D} $ the
effective coupling of the $c\overline{s} $ pair to the $D_{s} $  meson through
the Yukawa interaction $\lambda_{D}\overline{c}\gamma_{5}sD_{s} $, then the
amplitudes given by the two triangular loops can be written as :
\begin{equation}
Q_{2} + Q_{3} = \lambda_{D}(eg /2 \sqrt{2}) V_{cs} \varepsilon^{\mu}(P)
\overline{\varepsilon}^{\nu}(\it k) T_{\mu\nu}(P,\it k) \;.
\label{eq:3}
\end{equation}
where $T_{\mu\nu}$ has the following form :
\begin{equation}
T_{\mu\nu}(P,\it k) = K_{1} g_{\mu\nu} + K_{2} P_{\nu} \it k_{\mu} + K_{3}
P_{\mu} P_{\nu} + K_{4}\it k_{\mu} k_{\nu} + K_{5} P_{\mu} \it k_{\nu} + i
K_{6} \varepsilon_{\mu\nu\alpha\beta} P^{\alpha} \it k^{\beta}  \;.
\label{eq:4}
\end{equation}

In Eq.(4), the six terms $K_i(M_W^2, m_D^2, m_c, m_s), i = 1,6 $ are functions
of the external masses ($ P^2 = M_W^2, (P -\it k)^2 = m_D^2 $) as well as of
the internal fermions masses $m_{c}, m_{s}$ in the loops. The terms $K_{3},
K_{4} $,and $K_{5} $ do not contribute to the amplitude when they are
contracted with $\varepsilon^{\mu}(P) $ and  $\varepsilon^{\nu}(\it k) $, the
terms $K_{2} $ and $K_{6} $ are finite, while $K_{1} $ turns out to be
logarithmically divergent. However when the gauge-invariant condition is
imposed to the full amplitude $Q = Q_{1} + Q_{2} + Q_{3} $,( i.e. the amplitude
Q must vanish under the replacement $\varepsilon^{\nu}(\it k) $ by $\it k^{\nu}
$), then the $K_{1} $ term is completely fixed by $F_{D} $ and $K_{2} $, its
computation is therefore unnecessary. The constraint required by
gauge-invariance is given by :
\begin{equation}
 F_{D} + \lambda_{D} [ K_{1} + ( P\cdot \it k) K_{2} ] = \rm 0 \;.
%label{eq:5}
\end{equation}
This condition is reminiscent of the similar problem arises in the three
indices $R_{\mu\nu\tau} $ amplitude met long time ago by Rosenberg $^{[7]} $ in
the triangular loop integration : among the eight terms in the development of
$R_{\mu\nu\tau} $, there are two divergent ones that can be related by
gauge-invariance to other finite terms, their computations can therefore be
avoided. This condition has been employed in the Adler paper $^{[3]} $.

Putting together Eqs.(2), (3), (4) and (5), the full amplitude Q has a simple
form with only two finite terms $K_{2} $ and $K_{6} $ :$^{[1]} $
\begin{eqnarray}
Q & = & \lambda_{D} (eg / 2 \sqrt{2}) V_{cs} \varepsilon^{\mu}(P)
\overline{\varepsilon}^{\nu}(\it k) \{ [ P_{\nu} \it k_{\mu} -
g_{\mu\nu}(P\cdot \it k)] K_2 (M_W^2, m_D^2, m_c, m_s) \nonumber \\
     &  & \hspace{30mm} + i \varepsilon_{\mu\nu\alpha\beta} P^{\alpha} \it
k^{\beta} K_6 (M_W^2, m_D^2, m_c, m_s) \}  \;.
\label{eq.6}
\end{eqnarray}
The $K_{2} $ and $K_{6} $ are calculated to be :
\begin{eqnarray}
 K_2(M_W^2, m_D^2, m_c, m_s) = N [ Q_{c} A^{c}(P,\it k) + Q_{s} A^{s}(P,\it k)
] = 2 A^{c} - A^{s} ,    \nonumber\\
\hspace{3mm}
K_6(M_W^2, m_D^2, m_c, m_s) = N [ Q_{c} V^{c}(P,\it k) + Q_{s} V^{s}(P,\it k) ]
= 2 V^{c} - V^{s} \;.
\label{eq:7}
\end{eqnarray}
where N is the number of colors, $Q_{c,s} $ are respectively the charges( in
units of $|e|$ ) of the charm and strange quarks coupled to the photon, finally
the $A^{c,s}(P,\it k) $ and $V^{c,s}(P,\it k) $ terms are related to the
contributions of the axial-vector and vector-current of the quarks coupled to
the W boson.

They are given by :
\begin{equation}
A^{c}(M_W^2, m_D^2, m_c, m_s)  = \frac{1}{4\pi^{2}} \ \int_{0}^{1}\!\, dx
\int_{0}^{1 - x }\!\, dy  \frac{m_{c} - x (m_{c} + m_{s}) - 2 x y (m_{c} -
m_{s})}{M_{W}^{2} x (1 - x - y) + m_{D}^{2} x y - m_{c}^{2}(1 - x) - m_{s}^{2}
x } \   \;.
\label{eq:8}
\end{equation}

\begin{equation}
A^{s}(M_W^2, m_D^2, m_c, m_s) =  \frac{1}{4\pi^2}\ \int_{0}^{1}\!\, dx
\int_{0}^{1 - x }\!\, dy \frac {m_{c} + x (m_{c} + m_{s}) - 2 x y (m_{c} -
m_{s})}{M_{W}^{2} x (1 - x - y) + m_{D}^{2} x y - m_{s}^{2}(1 - x) -m_{c}^{2} x
}\   \;.
\label{eq:9}
\end{equation}

\begin{equation}
V^{c}(M_W^2, m_D^2, m_c, m_s)  = \frac{1}{4\pi^2}\ \int_{0}^{1} dx \int_{0}^{1
- x } dy \frac{m_{c} - x (m_{c} - m_{s})}{M_{W}^{2} x (1 - x - y) + m_{D}^{2} x
y - m_{c}^{2}(1 - x) -m_{s}^{2} x }\   \;.
\label{eq:10}
\end{equation}

\begin{displaymath}
V^{s}(M_W^2, m_D^2, m_c, m_s)  = \frac{1}{4\pi^2}\ \int_{0}^{1} dx \int_{0}^{1
- x } dy  \frac{m_{s} + x (m_{c} - m_{s})}{M_{W}^{2} x (1 - x - y) + m_{D}^{2}
x y - m_{s}^{2}(1 - x) -m_{c}^{2} x } \
\end{displaymath}
\begin{equation}
       =  V^{c}(M_W^2, m_D^2, m_s, m_c)      \;.
\label{eq:11}
\end{equation}
which can be reexpressed in terms of the integral representation :
\begin{equation}
A^{c,s} =  \frac {m_{c} - m_{s}}{M_{W}^{2} - m_{D}^{2}} \ + \frac
{F^{c,s}}{(M_{W}^{2} - m_{D}^{2})^{2}} \  \;.
\label{eq:12}
\end{equation}
\begin{equation}
V^{c,s} = {\pm} \frac {m_{c} - m_{s}}{M_{W}^{2} - m_{D}^{2}} \  G_{0}^{c,s} -
\frac {m_{c,s}}{M_{W}^{2} - m_{D}^{2}} G_{1}^{c,s} \  \;.
\label{eq:13}
\end{equation}
The $ + $ and $ - $ signs in Eq.(13) are associated to $V^{c} $ and $V^{s} $
respectively.
\begin{equation}
F^{c,s}(M_W^2, m_D^2, m_c, m_s) = \frac{1}{4\pi^2}\ \int_{0}^{1} dx  { \frac
{\alpha_{c,s}(x)}{x} \ } \log [1 - x (1 - x)\beta_{c,s}(x) ] \  \;.
%label{eq:14}
\end{equation}
\begin{equation}
G_{j}^{c,s}(M_W^2, m_D^2, m_c, m_s) = \frac{1}{4\pi^2}\ \int_{0}^{1} dx  \frac
{1}{x^{j}} \ \log [1 - x (1 - x) \beta_{c,s}(x) ] \  \hspace{2mm} (j = 0,1) \;.
%label{eq:15}
\end{equation}
where :
\begin{eqnarray}
\alpha_{c,s}(x) & = & 2 (m_{c} - m_{s})[M_{W}^{2} x (1 - x) - m_{c}^{2}(1 - x)
- m_{s}^{2} x ] \nonumber \\
                &   &  - (M_{W}^{2} - m_{D}^{2})[ m_{c,s}(1 {\mp} x) {\mp}
m_{s,c} x ] \;.
%label{eq:16}
\end{eqnarray}

\begin{equation}
\beta_{c,s}(x) =  \frac {M_{W}^{2} - m_{D}^{2}}{M_{W}^{2} x (1 - x) -
m_{c,s}^{2}(1 - x) - m_{s,c}^{2} x } \ \;.
\label{eq:17}
\end{equation}
In Eq.(16), the $- $ and $+ $ signs correspond respectively to the first and
the second lower index i, j of $m_{i,j} $. In Eq.(6), the three couplings
$\lambda_{D}, e, gV_{cs} $ associated respectively to $D_{s}, \gamma $ and W
are explicitly factorized out, all the dynamics are contained in the $K_{2} $
and $K_{6} $ quantities through Eqs.(7)...(11) that reflect the one loop
integration in which one recognizes the $\frac{1}{4 \pi^{2}} $ factor together
with the internal quarks masses $m_{c} $, $m_{s} $ as well as the external ones
$M_{W}^{2}, m_{D}^{2} $. Squaring the amplitude Eq.(6), summing and averaging
over the polarizations of the photon and the W, carring out the phase-space
integrations, one gets :
\begin{equation}
\Gamma(W^{\pm} \rightarrow \gamma + D_{s}^{\pm}) = \lambda_{D}^{2} \frac {e^{2}
g^{2} |V_{cs}|^{2}}{768 \pi} \ M_{W}^{3} (1 - \frac {m_{D}^{2}}{M_{W}^{2}} \
)^{3} (|K_{2}|^{2} + |K_{6}|^{2}) \;.
\label{eq:18}
\end{equation}

We remark that $F^{c,s} $ and $G_{j}^{c,s} $ and consequently $K_{2} $ and
$K_{6} $ are complex functions because of the logarithm in the integrands : the
complexity of these functions can be easily understood since the energy $M_{W}
$ is far beyond the threshold $m_{c} + m_{s} $, then the W boson can first
decay into a real $c + \overline{s} $ pair which are converted afterwards to
the final hadronic states, therefore the triangle graph develops an imaginary
part.
Numerically we get, using $m_{c} = 1.5 $ GeV, $m_{s} = 0.2 $ GeV  :
\begin{eqnarray}
K_{2}(P,\it k) = (1.56 - 1.92 i) \hspace{3mm} 10^{-4}  GeV^{-1}  \nonumber\\
\hspace{3mm} K_{6}(P,\it k) = (2.20 - 2.22 i) \hspace{3mm} 10^{-4}  GeV^{-1}
\;.
\label{eq:19}
\end{eqnarray}
These numbers are rather insensitive to the variation of $m_{c} $ taken between
1.3 - 1.7 GeV, $m_{s} $ between 0.15 - 0.5 GeV. The only unknown parameter for
the width in Eq.(18) is the dimensionless effective coupling $\lambda_{D} $ to
which is devoted the following section.

\vspace{10mm}
\hspace{30mm}................................................................\\

\large
The cases of  $\pi^{o} \rightarrow \gamma \gamma $ ,$Z^{0} \rightarrow \pi^{0}
+ \gamma $,$W^{+} \rightarrow \gamma + \pi^{+} $ and the effective
hadron-quarks coupling. \\

\normalsize
We first reconsider the classical $\pi^{o} \rightarrow \gamma \gamma $ case as
an illustration for the determination of the effective pseudoscalar meson
quark-antiquark coupling.
The triangle diagrams we employ to calculate the decay amplitude $W^{\pm}
\rightarrow \gamma + D_{s}^{\pm} $ are nothing else but a copy of the famous
$\pi^{o} \rightarrow \gamma \gamma $ ones computed long time ago by Steinberger
$^{[8]} $ in which the fermion circulating in the loop is the proton and
$\lambda $  is the strong pion-nucleon coupling $g_{\overline{N}N\pi^{o}} $
which is related to the proton mass $M_{N} $ and the pion decay constant
$F_{\pi} $  via the Goldberger-Treiman relation  $g_{\overline{N}N\pi^{o}} =
\frac {\sqrt{2} M_{N}}{F_{\pi}} \ $. Today, in the modern QCD language, the
proton is naturally replaced by the up-down quarks and the pion-nucleon
coupling   $g_{\overline{N}N\pi^{o}} $ becomes an effective coupling between
the $\pi^{o} $ and the quarks involved in the triangle diagrams :
\begin{equation}
\lambda_{u} \equiv g_{\overline{u}u\pi^{o}} = \frac {\sqrt{2} m_{u}}{F_{\pi}} \
,\hspace{10mm} \lambda_{d} \equiv g_{\overline{d}d\pi^{o}} = {-}\hspace{3mm}
\frac {\sqrt{2} m_{d}}{F_{\pi}} \ \;.
\label{eq:20}
\end{equation}
where $m_{u,d} = m  $ is the up, down quark mass taken to be equal for
simplicity. Since only the vector currents ( and not the axial-vector ones )
are relevant to photons in  $\pi^{0} \rightarrow \gamma \gamma $, its amplitude
analogous to our Eq.(6) is given by the vector part $V^{q} $ of the quarks.
\begin{equation}
Q(\pi^{0} \rightarrow \gamma \gamma ) = i e^{2} \overline{\varepsilon}^{\mu}(P)
 \overline{\varepsilon}^{\nu}(\it k) \varepsilon_{\mu\nu\alpha\beta}
P^{\alpha}\it k^{\beta} \{ \sum_{q=u,d}^{} (\lambda_{q} Q_{q}^{2} V^{q}) N \}
\;.
\label{eq:21}
\end{equation}
where $V^{q} $ is twice the corresponding $V^{c,s} $ of our Eqs.(10),(11) with
appropriate modifications of the arguments: ($M_{W} \rightarrow 0, m_{D}
\rightarrow m_{\pi},\hspace{1mm} m_{c,s} \rightarrow m_{u,d} = m $). The origin
of the factor two is that for each triangle graph, there must be added a second
one in which the two photons are crossed due to the Bose statistic. If the
pion, as a Goldstone boson
, is taken to be massless, then Eq.(10) or (11), after appropriate replacements
of the arguments as given above, becomes :
\begin{displaymath}
V(0,0,m,m) = - \frac {1}{8 \pi^{2}} \ \frac {1}{m} \
\end{displaymath}
such that
\begin{equation}
V^{u}  = V^{d} =  - \frac {1}{4 \pi^{2}} \ \frac {1}{m} \ \;.
\label{eq:22}
\end{equation}

Putting Eqs.(20) and (22) into Eq.(21), we get :
\begin{equation}
Q(\pi^{0} \rightarrow \gamma \gamma ) = - i e^{2}
\overline{\varepsilon}^{\mu}(P)  \overline{\varepsilon}^{\nu}(\it k)
\varepsilon_{\mu\nu\alpha\beta} P^{\alpha}\it k^{\beta} \frac{\sqrt{2}}{4 \pi^2
F_{\pi}} \ \;.
\label{eq:23}
\end{equation}
in perfect agreement with data, justifying a posteriori, the correctness of the
effective coupling $\lambda_{q} $ as given by the extended G-T relation i.e.
Eq.(20).\\

Two remarks are in order :\\

(i). The amplitude Eq.(23) is independent on m, the mass of the internal
fermion circulating in the loop. This property is only a particular case due to
the fact that the masses of the external fields are neglected. Indeed, the m
dependence of V(0,0,m,m) as given by the denominator in Eq.(22) is exactly
cancelled by the same m dependence in the numerator of the coupling
$\lambda_{q} $ via Eq.(20).
This property, which is also true in the case of $W^{+} \rightarrow \gamma + $
PseudoGoldstone of technicolor model ( because all the external masses are
negligible compared to the techniquark masses in the loop ) $^{[F2]} $, no
longer holds when the external masses are dominant as in the cases $W^{\pm}
\rightarrow \gamma + \pi^{\pm} $, $W^{\pm} \rightarrow \gamma + D_{s}^{\pm} $
that will be discussed later. This property is  also summarized in Table 1.\\

(ii). The effective coupling $\lambda_{q} $ as given by the G-T relation
Eq.(20) can also be derived$^{[6]} $ in the Nambu, Jona-Lasinio theory, $^{[5]}
$ a prototype which dynamically generates masses for both fermions and bosons.
\\

We now go further and consider the cases $Z^{0} \rightarrow \pi^{o} + \gamma $
, $W^{\pm} \rightarrow \gamma + \pi^{\pm} $ that have been thoroughly analysed
by AMP. In some sense, these modes are the inverse of the $\pi^{0} \rightarrow
\gamma \gamma $ case, in which one photon $\gamma $ is off-mass shell. Consider
first the $W^{\pm} \rightarrow \gamma + \pi^{\pm} $, the case $Z^{0}
\rightarrow \pi^{0} + \gamma $ is similar.$^{[9]} $  \\
\hspace{30mm}.................................................................\\ \small \em
\hspace{3mm}[F 2]. In the last part of the AMP paper, the authors implicitly
use the scheme we advocate here ( triangular diagrams + effective hadron quark
coupling ) to calculate $W^{\pm} \rightarrow \gamma + Technipion. $  \\

\normalsize \rm
The $W^{\pm} \rightarrow \gamma + \pi^{\pm} $ amplitude is again given by
Eq.(6) with appropriate substitution in the coupling $\lambda $ and in the
arguments of $K_{2}, K_{6} $ :
\begin{equation}
\lambda_{D} \rightarrow \lambda_{\pi} \equiv g_{\pi^{+}u\overline{d}} =
\sqrt{2} g_{\pi^{o} u \overline{u}} = \frac {2 m}{F_{\pi}} \  \;.
\label{eq:24}
\end{equation}

Also the internal fermion masses $m_{c,s} $ in Eq(8)...(11) are replaced by the
common up-down quark mass m, such that the corresponding $K_{2}, K_{6} $ for
the
$W^{\pm} \rightarrow \gamma + \pi^{\pm} $ case, after appropriate substitution
mentioned above, can be explicitly calculated and put in an analytic form :
\begin{equation}
K_{2}(P^{2}=M_{W}^{2}, m_{\pi}^{2}=0, m, m) = \frac {1}{4 \pi^{2}} \ \frac
{m}{M_{W}^{2}} \ [ \frac {L^{2}}{2} \ - 6 \beta L + 12 ]  \;.
\label{eq:25}
\end{equation}

\begin{displaymath}
K_{6}(P^{2}=M_{W}^{2}, m_{\pi}^{2}=0, m, m) = \frac {1}{4 \pi^{2}} \ \frac
{m}{M_{W}^{2}} \  \frac {L^{2}}{2} \
\end{displaymath}
where
\begin{displaymath}
L(M_{W}^{2}, m^{2}) = \log \frac {1 + \beta}{1 - \beta} \ - i {\pi}
\hspace{5mm}
and \hspace{10mm} \beta = \sqrt{1 - \frac {4 m^{2}} { M_{W}^{2} }\ }
\end{displaymath}

\hspace{32mm} $\simeq \log ( \frac {M_{W}^{2}}{m^{2}}\ ) - i {\pi} $
\hspace{5mm} ( for $M_{W}^{2} \gg m^{2} $ ) \\

The presence of the imaginary part is already discussed. Putting together
Eqs(24) and (25) into Eq.(6) we get for the $W^{\pm} \rightarrow \gamma +
\pi^{\pm} $ amplitude :
\begin{eqnarray}
Q(W^{\pm} \rightarrow \gamma + \pi^{\pm}) & = & (e g /2\sqrt{2}) V_{ud}
\varepsilon^{\mu}(P) \overline{\varepsilon}^{\nu}(\it k) \cdot  \nonumber \\
&  & \{[P_{\nu}\it k_{\mu} -g_{\mu\nu}(P \cdot \it k)] \overline{K_{2}} + i
\varepsilon_{\mu\nu\alpha\beta} P^{\alpha} \it k^{\beta} \overline{K_{6}} \}
\;.
\label{eq:26}
\end{eqnarray}
where
\begin{equation}
\overline{K_{2}} \equiv \lambda_{\pi} K_{2} = \frac{1}{4\pi^{2}} \ \frac
{1}{F_{\pi}} \ \frac {m^{2}}{M_{W}^{2}} \ [24 - 12 \eta + \eta^{2}] \;.
\label{eq:27}
\end{equation}

\begin{equation}
\overline{K_{6}} \equiv \lambda_{\pi} K_{6} = \frac{1}{4\pi^{2}} \ \frac
{1}{F_{\pi}} \ \frac {m^{2}}{M_{W}^{2}}  \eta^{2}
\hspace{10mm} (with \hspace{3mm} \eta = \log \frac{M_{W}^{2}}{m^{2}} \ - i
{\pi} ) \;.
\label{eq:28}
\end{equation}

Absorbing the effective coupling $\lambda_{\pi} $ into the dynamical parts
$K_{2}, K_{6}, (\overline{K_{i}} = \lambda_{\pi}{K_{i}})  $ we are now in a
position to compare our Eq.(26) with the AMP paper, namely their Eq.(2.11).
These authors do not use, neither the triangle graphs nor the effective
hadron-quarks coupling as we do in this paper, but instead they assume the
Brodsky-Lepage (BL) asymptotic off-shell photon $\pi^{o}\gamma\gamma $ form
factor $^{[10]} $ calculated in QCD. In our notation, they simply assume :

\begin{equation}
\overline{K_{6}} = \frac{F_{\pi}}{M_{W}^{2}} \;.
\label{eq:29}
\end{equation}
where the right-hand side of Eq.(29) $\frac{F_{\pi}}{P^{2}} \ $ is the
asymptotic $(P^{2} \rightarrow \infty) $ BL form factor. Also AMP assume that
asymptotically $\overline{K_{2}} = \overline{K_{6}} $. For the mass scale in
question ,i.e. when $P^{2} = M_{W}^{2} $ is very large, our Eq.(28) for
$\overline{K_{6}} $
is in agreement asymptotically with the BL form factor
$\frac{F_{\pi}}{M_{W}^{2}} $. The work of BL suggests that as $P^{2}
\rightarrow \infty $, the form factor $\overline{K_{6}} $ decreases with a
power law behaviour $(P^{2})^{-n} $ with n = 1, and the residue is fixed by the
scale $F_{\pi} $ representing the non-pertubative QCD dynamics of the pion.
In our model, Eq.(28) indicates that at large $P^{2} $ the form factor
$\overline{K_{6}} $  also behaves like $(P^{2})^{-1} $, however the residue,
obtained from the triangle loop integral, is equal to $\frac{1}{4 \pi^{2}} \
\frac{m^{2}}{F_{\pi}} \ [(\log \frac {P^{2}}{m^{2}})^{2} + \pi^{2} ] $.
Here, we take the absolute value of $|K_{6}| $ since it is complex, instead of
a real number $F_{\pi} $ in the BL work.
The residue $\frac{1}{4 \pi^{2}} \ \frac{m^{2}}{F_{\pi}} \ [(\log \frac
{P^{2}}{m^{2}})^{2} + \pi^{2} ] $ tells us that the dynamic in $W^{\pm}
\rightarrow \gamma + \pi^{\pm} $
still remembers ( through the m dependence ) the quarks bound inside the pion.
Moreover if we try to identify our residue with the Brodsky-Lepage one (i.e.
$F_{\pi} $), then we get an equation that determines the mass m :
\begin{equation}
\frac{1}{4 \pi^{2}} \ \frac{m^{2}}{F_{\pi}} \ [(\log \frac {P^{2}}{m^{2}})^{2}
+ \pi^{2} ] = F_{\pi} \;.
\label{eq:30}
\end{equation}
or equivalently :
\begin{equation}
4 \pi^{2} x = (\log x \ )^{2} + 2 (\log \frac {P^{2}}{F_{\pi}^{2}} \ ) \log x +
\pi^{2} + (\log \frac {P^{2}}{F_{\pi}^{2}} \ )^{2} \;.
\label{eq:31}
\end{equation}
with $ x = \frac{F_{\pi}^{2}}{m^{2}} \ $ \\

This non-linear equation has an unique solution $m \simeq 55 $ MeV which could
be considered as the running mass of the up-down quark at the scale $M_{W} $:
$m(\mu = M_{W}) \simeq 55 $ MeV .\\

It is a good surprise that our model, when matched with the BL one, gives a
plausible value $m = 55 $ MeV to the up-down quark mass. In other words, our
amplitude $W^{\pm} \rightarrow \gamma + \pi^{\pm} $ as given by Eqs.(26)..(28)
is the same as the AMP one if we take $m = 55 $ MeV . For such value of m, the
effective coupling $\lambda_{\pi} $ as given by Eq.(24) is equal to 0.83. Also
the numerical values of the dynamical quantities $K_{2,6} $ given by Eq.(25)
become :
\begin{eqnarray}
K_{2} = (0.56 - 0.58 i) \hspace{3mm} 10^{-5}  GeV^{-1}  \nonumber\\
K_{6} = (2.16 - 0.99 i) \hspace{3mm} 10^{-5}  GeV^{-1}  \;.
\label{eq:32}
\end{eqnarray}
These values for $W^{\pm} \rightarrow \gamma + \pi^{\pm} $ are to be compared
with the numerical ones of the $W^{\pm} \rightarrow \gamma + D_{s}^{\pm} $ case
given in Eq.(19).\\

Encouraged by the above results where the use of the effective pion
quark-antiquark $\lambda_{\pi} $ coupling ( together with the triangle diagram
) turns out to be satisfactory in $\pi^{o} \rightarrow \gamma \gamma $, as well
as in $W^{\pm} \rightarrow \gamma + \pi^{\pm} $, $Z^{0} \rightarrow  \gamma +
\pi^{o} $ for which experimental limits are known$^{[11]} $,
we come back to our decay
$W^{\pm} \rightarrow \gamma + D_{s}^{\pm} $ in Eq.(6). The dynamical parts
symbolized by $K_{2,6} $ have been computed and their numerical values are
given in Eq.(19), it remains the coupling $\lambda_{D} $.

We conjecture that $\lambda_{D} $ is of the same order as $\lambda_{\pi} = 0.83
$ calculated above. Indeed, in the old days of the strong interaction with
SU(3) flavour symmetry, all the couplings : pion-nucleon, Kaon-hyperons  etc...
as compiled in Ref[12] turn out to be similar , the inclusion of charm with
SU(4) symmetry seems plausible.\\

It is an ascertainment as a fact that the strong couplings are similar in the
flavour world of particles, it is their masses and their weak interaction
properties that differentiate them. \\

Then let us assume $\lambda_{D} \simeq \lambda_{\pi} $, in such case we obtain
$B(W^{\pm} \rightarrow \gamma + D_{s}^{\pm}) \simeq 10^{-6} $.
On the other hand, if we try the extreme case in which the $D_{s} $ is
considered as a Goldstone boson on the same footing as the pion ( at the W mass
scale the $D_{s} $ is still very light ), then the G-T relation extended to
charm would be $\lambda_{D} = \frac{m_{c} + m_{s}}{F_{D}} \simeq 8 $. This
extreme case which is not firmly justified, leads to an unplausible large value
$5\cdot 10^{-5} $ for the branching ratio.\\

Another guess $$\frac{\lambda_{D}}{\lambda_{\pi}} = \frac{(m_{c} +
m_{s})/F_{D}}{(\overline{m}_{u} + \overline{m}_{d})/F_{\pi}} $$ where
$\overline{m}_{u,d} $ are the constituent$^{[F3]} $ up-down quark mass $\simeq
$ 300 MeV - would give a branching ratio $ \simeq 2 \cdot 10^{-6} $, which is
also our preferred value.\\

Compared to the AMP work, the main difference with ours rests in the dynamical
quantities $K_{2}, K_{6} $. By assuming the BL asymptotic form-factors, AMP
argue that all radiative decay modes of $W^{\pm} $  into a pseudoscalar meson
$P^{\pm} $ : $\pi^{\pm}, K^{\pm}, D^{\pm}, D_{s}^{\pm}, B^{\pm} $ are
approximately equal(beside the Cabibbo-Kobayashi-Maskawa CKM mixing ). Since
their form-factors $\overline{K}_{2,6} $, as fixed by $F_{\pi}, F_{K}, F_{D},
F_{B} $  are similar, the branching ratios turn out to be $3\cdot 10^{-9} $ for
$\pi^{\pm} $ and $D_{s}^{\pm} $ and much smaller for $K^{\pm} $,
$D^{\pm} $, $B^{\pm} $ by CKM suppression.

On the contrary, in our model, the dynamical terms $K_{2},K_{6} $(given by
triangular loop integral) still remember the quarks - via their masses - inside
the $P^{\pm} $.
Therefore in our scheme, the rate $W^{\pm} \rightarrow \gamma + D_{s}^{\pm} $
is much larger than the $W^{\pm} \rightarrow \gamma + \pi^{\pm} $ one by at
least a factor $({m_{c}}/{m_{u}})^{2}$. ( See Table 1 and our Eqs (19),(31)).\\
\hspace{10mm}................................................................\\
\small \em [F3]. It is quite possible that the BL form factors - employed by
AMP in the calculation of the $W^{\pm} \rightarrow \gamma + \pi^{\pm} $ mode -
underestimate its rate, since it corresponds to the use of a value 55 MeV (
quasi current mass ) for the up down quark. If instead we take the constituent
mass $\simeq 300 $ MeV , then  the $W^{\pm} \rightarrow \gamma + \pi^{\pm} $
branching ratio would be $10^{-7} $(instead of $3\cdot 10^{-9} $ in the AMP
calculation)

\normalsize \rm

In the AMP scheme, the effective coupling $\lambda_{P} $ must decrease with the
quark-mass like $\frac {1}{m_{q}} \ $ in order to compensate the dependence
like $m_{q} $ of $K_{2},K_{6} $, such that the amplitudes $\overline{K}_{2,6} =
\lambda_{q} K_{2,6} $ are independent on $m_{q} $. We don't understand why the
effective coupling of K meson with strange quark, D meson with charm quark and
B meson with b quark is smaller and smaller like $\frac{1}{m_{q}} $, while the
ones of pion with up-down quark or technipion
with techniquark, as given by the G-T relation,are proportional to $m_{q} $. \\

In conclusion, we present in details the reasons  - summarized in Table 1 - why
the mode $W^{\pm} \rightarrow \gamma + D_{s}^{\pm} $ is strongly enhanced
compared to $W^{\pm} \rightarrow \gamma + \pi^{\pm} $ , and could be observable
with a branching ratio around $10^{-6} $. But the last words, as always,are on
the experimental side.  Only experiments can tell us whether or not the mode
discussed here has a viable chance of being detected, and we are eager to learn
from experiments the possible observation  of photon and $D_{s} $ in the W
decay. The interests for this mode are mentioned from the beginning.

\bigskip
\vspace{180mm}

\noindent
\begin{center}
\large \rm References \\
\end{center}

\normalsize

[1]. L.Arnellos, W. J. Marciano and Z. Parsa, Nucl. Phys. B196, 378 (1982) \\

[2]. Y. Y. Keum and X. Y. Pham in preparation (1993) \\

[3]. S. Adler, Phys. Rev. 177, 2426 (1969) \\

[4]. J. S. Bell and R. Jackiw, Nuovo Cimento 60A, 37 (1969) \\

[5]. Y. Nambu and G. Jona-Lasinio, Phys. Rev. 122, 345 (1961) \\

[6]. S.P. Klevansky, Rev. Mod. Phys. 64, 649 (1992) \\

[7]. L. Rosenberg, Phys. Rev. 129, 2786 (1963)  \\

[8]. J. Steinberger, Phys. Rev. 76, 1180 (1949)  \\

[9]. T. N. Pham and X. Y. Pham, Phys. Lett. 247B, 438 (1990) \\

[10]. G. P. Lepage and S. J. Brodsky Phys.Lett. B87, 359 (1979) \\

\hspace{55mm} and Phys. Rev. D22, 2159 (1980) \\

[11]. Review of Particle properties, PDG, Phys. Rev. D45, 11II (1992) \\

[12]. O. Dumbrajs et al, Nucl. Phys. B216, 277 (1983)  \\

\bigskip

\vspace{150mm}

\begin{tabular}{|p{1.35in}|p{1.5in}|p{1.25in}|p{1.35in}|}
\hline
\multicolumn{4}{|c|}{Table. I}
  \\  \hline\hline

$$ \pi^{o} \rightarrow \gamma  + \gamma $$ & $$W^{\pm} \rightarrow \gamma +
P^{\pm}_{Tech.} $$ & $$W^{\pm} \rightarrow \gamma + \pi^{\pm} $$ & $$W^{\pm}
\rightarrow \gamma + D_{s}^{\pm} $$ \\ \hline

$$ \lambda_{\pi} \sim \frac{m_{u}}{F_{\pi}}\ $$ & $$ \lambda_{P_{tech}} \sim
\frac{M_{Techniquark}}{F_{Technipion}} \ $$ & $$ \lambda_{\pi} \sim
\frac{m_{u}}{F_{\pi}}\ $$ & $$ \lambda_{D} \geq \lambda_{\pi} $$ \\ \hline

$$K_{2,6} \sim \frac{1}{m_{u}} \ $$ & $$K_{2,6} \sim \frac{1}{M_{Techniquark}}
\ $$ & $$K_{2,6} \sim \frac{m_{u}}{M_{W}^{2}} \ $$ & $$K_{2,6} \sim
\frac{m_{c}}{M_{W}^{2}} \ $$ \\ \hline

$$ \overline{K}_{2,6} \equiv \lambda K_{2,6}
                      \sim \frac{1}{F_{\pi}} \  $$ & $$ \overline{K}_{2,6} \sim
\frac{1}{F_{Technipion}} $$ & $$ \overline{K}_{2,6} \sim \frac
{m_{u}^{2}}{M_{W} ^{2}}\  \frac{1} {F_{\pi}} \ $$ & $$ \overline{K}_{2,6} \geq
(\frac{m_{c}m_{u}}{M_{W}^{2}})\frac{1}{F_{\pi}} \ $$ \\  \hline
\end{tabular}

\vspace{30mm}

Figures captions : Feynman diagrams for the process $W^{+} \rightarrow \gamma +
D_{s}^{+} $.

\vspace{30mm}

Table 1 caption : Treatment of the four decays $\pi^{o} \rightarrow \gamma
\gamma, \hspace{2mm}
W^{\pm} \rightarrow \gamma  P^{\pm}_{Technipion} $,\\

\hspace{30mm} $W^{\pm} \rightarrow \gamma  \pi^{\pm},\hspace{2mm} W^{\pm}
\rightarrow \gamma  D_{s}^{\pm} $ \hspace{3mm} by a common approach.\\

\end{document}